# The Arrow of Time

Ernst Karl Kunst
Im Spicher Garten 5
53639 Königswinter
Germany
e-mail: ErnstKunst@aol.com

**It is shown that (special) relativistically dilated time is the vector sum of rest time and time induced by movement in three dimensional space exeeding the rest time component and that the first vector is orthogonally directed relative to our three dimensional space. This again implies that its origin lies in the movement of three dimensional space relative to a four dimensional manifold. The theory predicts asymmetrical time dilation**.

It has been shown in [1] that translational velocity is symmetrically composite and the resulting space-time geometry Gaussian. Accordingly the Lorentz factor $\gamma_0 = (1 - v_0^2/c^2)^{-1/2}$ associated with the composite velocity $v_0$ is the complex number $n(\beta_0', 1)$ in a complex ct, x-plane of space-time

$$n = \gamma_0 (\cos\Phi + i\sin\Phi) \qquad (1)$$

so that

$$|n| = \sqrt{\gamma_0^2} = \gamma_0 \qquad (1a)$$

($\beta_0' = v_0 dt'/(cdt)$ and $dt' = dt\gamma_0$). Furthermore, in the same study has been demonstrated Einstein's "relativity of simultaneity" and the "FitzGerald-Lorentz contraction" be erroneous derivations of the Lorentz transformation. Instead the correct interpretation of the latter predicts a relativistic expansion of length or volume by the factor $\gamma_0$ and the simultaneity of events, but dilated by the same factor. The well-known rise of interaction-radii of hadrons in high energetic collisions and related experimental phenomena are due to this relativistic expansion of volume - as has been extensively shown. Therefrom directly follows $m_t = dt_x' v_0/c = dt_x \gamma_0 v_0/c$, where $dt_x = dx/c$, which means that relativistic mass is solely induced by the factor $dt_x' v_0/c$ of the expanded volume $V' = dx'dy'dz'$ of a moving body and, furthermore, the existence of a fundamental length $\lambda_0 = c\tau_0 = \sqrt{h}$ and a quantum of time $\tau_0$.

Because the space-like vector **$dt'_x v_0/c$** and rest time $dt_x$ are of the same origin [2] the latter must beyond its character as a "time-like vector" be a genuine, directionally well defined vector ($dt_x$), too, implying relativistically dilated time to be the vector sum:

$$dt_x + \frac{dt'_x v_0}{c} = dt'_x \qquad (2)$$

or, especially,

$$T_0 + \left|\frac{T_0}{dt_x}\right| \times \frac{dt'_x v_0}{c} = T'_0. \qquad (2a)$$

Thus, time in the moving frame is the vector sum of the rest time in the frame of the observer and the space-like vector $dt'_x v_0/c$ directed into any of the dimensions of three dimensional space. From (1) follows that $dt'_x v_0/c$ is based orthogonally onto the vector $dt_x$ ($dt'_x v_0/c \perp dt_x$) so that the absolute value of modulus of the vector $dt'_x$ and the vector $T_0'$, respectively, is

$$|dt'_x| = \frac{dt_x}{\sqrt{1 - \frac{v_0^2}{c^2}}},$$

$$|T'_0| = \frac{T_0}{\sqrt{1 - \frac{v_0^2}{c^2}}}. \qquad (3)$$

From the foregoing is evident that any observer can consider time in his own rest frame as a vector sum $dt_{xmin} + dt_x v_0/c = dt_x$ ($T_{0min} + |T_{0min}|/|dt_{xmin}| \times dt_x v_0/c = T_0$) owing to the movement relative to a system where (3) attains the minimum value $|dt_x| = dt_{xmin}$ and $|T_0| = T_{0min}$, respectively, which presumably will be the case if the latter frame is resting relative to the cosmic microwave background (space).

It is clear that the vector $dt_{xmin}$ ($T_{0min}$) is not a vector sum and remains invariant independent on the sense of the vector $dt_x v_0/c$ associated with the movement of any rest frame in three dimensional space. Thus, the direction of the vector $dt_{xmin}$ ($T_{0min}$) must show into a fourth geometrical dimension outside of three dimensional space. This result already has been infered from the equivalence of rest time and dilated time [2].

Let us assume that the absolute value of modulus according to (3) is really associated with a higher velocity of the moving frame relative to space and, therewith, to the observer's rest frame so that is valid

$$\int_x dt_x = \int_{x'} dt'_x \sqrt{1 - \frac{v_0^2}{c^2}}.$$

Because of the absolute symmetry of both frames according to the principle of relativity a Lorentz transformation in either frame ($dt'_x = dt_x \gamma_0$ and $dt_x = dt'_x \gamma_0$) results in:

$$\int_{x'} dt'_x = \int_x dt_x \sqrt{1 - \frac{v_0^2}{c^2}}.$$

This result is in accord with the well-known flight-time experiments with airplanes [4] and muons [5], [6]. Therefore, the twin paradox of special relativity is resolved to the result that time dilation is asymmetrically dependent on velocity relative to the microwave background (this result has been independently derived in [3]).

In conventional special relativity the above result is achieved by an one-sided Lorentz transformation from the moving to the resting frame. But that theory predicts the same result if the observer's rest frame is assumed to be the moving one - contrary to experiment. Although this is frequently denied is in the framework of conventional special relativity another conclusion absolutely not possible because it only admits of relative and denies absolute motion. Thus, to be in accord with experimentally verified asymmetric time dilation the conventional interpretation in reality tacitly involves absolute motion.

In the case of collision experiments a contradiction to special relativity does not arise, because the quantized inertial motion between any two frames of reference is applied to a preferred natural rest frame $\Sigma_0$ implying their absolute symmetry and equality relative to $\Sigma_0$ [1] - which again is moving relative to space (micro wave background). This will be deduced by every observer in every frame. On the other hand the different velocity of the frames under consideration relative to space and, therewith, different absolute values of modulus of **dt'$_x$** and **τ$_0$'** acccording to (3) must be real. Hence, the latter vectors always result from two successive Lorentz transformations so that in the case of parallel and antiparallel motion, respectively, we receive:

$$dt_{x_{min}} + \frac{dt_x u_0}{c} \pm \frac{dt'_x v_0}{c} = \pm dt'_x$$

and

$$T_{0_{min}} + \left|\frac{T_{0_{min}}}{dt_{x_{min}}}\right| \times \frac{dt_x u_0}{c} \pm \left|\frac{T_0}{dt_x}\right| \times \frac{dt'_x v_0}{c} = \pm T'_0,$$

where $u_0$ means the velocity of the rest frame of the observer relative to space and $v_0$ between the frames under consideration, respectively. This implies that the principle of relativity and, therewith, the group character of the Lorentz transformation remain valid, but with the restriction that of any two frames of reference a time interval dt of the frame with the higher velocity relative to space is really longer by the factor (1a) as compared with the other system.

The above formulas predict that all clocks in the solar system run slower than a clock resting relative to the microwave background, owing to its motion through the latter. Furthermore, all clocks on Earth must run slower if during the year her velocity is in the same direction spatially as the general motion of the sun and faster if her velocity is in the other direction - compared with a clock resting relative to the sun.

## References

[1] Kunst, E. K.: Is the Kinematics of Special Relativity incomplete?, physics/9909059
[2] Kunst, E. K.: On the Origin of Time, physics/9910024
[3] Kunst, E. K.: Is the Lorentz Transformation Distant-Dependent? physics/9911022
[4] Hafele, J.,Keating, R., Science 177, 166 (1972)
[5] Rossi, B., Hall, D., Phys. Rev. 59, 223 (1941)
[6] Farley, F. M., et al., Nature 217, 17 (1968)